\documentclass[prl,showpacs,twocolumn,amsmath,amssymb,superscriptaddress]{revtex4}
\usepackage{graphicx}
\usepackage{dcolumn}
\usepackage{bm}

\usepackage{graphicx}
\usepackage{dcolumn}
\usepackage{bm}

\usepackage{amsmath,amssymb}
\newcommand{\eps}{\varepsilon}
\def\mb{\mathbf}
\def\be{\begin{equation}}
\def\ee{\end{equation}}
\def\ba{\begin{eqnarray}}
\def\ea{\end{eqnarray}}
\usepackage{color}

\newcommand{\de}[1]{\textcolor{black}{#1}}

\newcommand{\bk}{{{\bf{k}}}}

\newcommand{\nn}{\nonumber}

\newcommand{\bea}{\begin{eqnarray}}
\newcommand{\eea}{\end{eqnarray}}

\begin{document}

\title{Electronic correlations stabilizing time-reversal broken chiral superconductivity in single-trilayer TiSe$_2$}

\date{Version 4, 18 Nov 2013, compiled \today}

\author{R. Ganesh}
\affiliation{Institute for Theoretical Solid State Physics, IFW-Dresden, D-01171 Dresden, Germany }
\author{G. Baskaran}
\affiliation{The Institute of Mathematical Sciences, C.I.T. Campus, Chennai 600 113, India}
\affiliation{Perimeter Institute for Theoretical Physics, Waterloo, Ontario, Canada N2L 2Y5}
\author{Jeroen van den  Brink}
\affiliation{Institute for Theoretical Solid State Physics, IFW-Dresden, D-01171 Dresden, Germany }
\affiliation{Department of Physics, TU Dresden, D-01062 Dresden, Germany}
\author{Dmitry V. Efremov}
\affiliation{Institute for Theoretical Solid State Physics, IFW-Dresden, D-01171 Dresden, Germany }

 \date{\today}

\begin{abstract}
Bulk TiSe$_2$ is an intrinsically layered transition metal dichalcogenide (TMD) hosting both superconducting and charge density wave (CDW) ordering. Motivated by the recent progress in preparing two-dimensional TMDs, we study these frustrated orderings in {\it single} trilayer of TiSe$_2$ within a renormalization group approach. We establish that a novel state with time-reversal symmetry broken chiral superconductivity can emerge from the strong competition between CDW formation and superconductivity. Its stability depends on the precise strength and screening of the electron-electron interactions in two-dimensional TiSe$_2$.
\end{abstract}



\maketitle

\paragraph{Introduction}
Transition metal dichalcogenides (TMDs) with the chemical formula MX$_2$, where M is a transition metal from groups IV-VI (Ti, Zr, Hf, V, Nb, Ta {\it etc.}) and X is a chalcogen element (Se, S, Te), are emerging as a new class of two-dimensional materials  with high potential for \de{nanoelectronics}
applications~\cite{Xu2013,Radisavljevic2013,HumbertoTerrones2013,wang2012}. The intense research activity in this field is inspired by the graphene boom, which was sparked by the possibility of manufacturing a purely two-dimensional material with high carrier mobility. TMDs consist of stacked X-M-X trilayers which, just as graphene, have hexagonal symmetry. These trilayers are held together by weak van der Waals forces, which allows exfoliation of the individual trilayers and the deposition of these layers onto various substrates~\cite{Coleman2011}.

Interestingly, the plethora of phenomena that occur in TMDs is even more multifarious than in graphene. Metallic TMDs not only have a generic instability towards the formation of different types of charge density waves (CDWs), but some of them also host superconductivity (SC). Moreover, due to the presence of transition metal elements electron-electron interactions can play a significant role. 
From the point of view of superconductivity, this is a highly interesting mix of ingredients. It is well known that the competition of SC with density-waves in the presence of electronic correlations may lead to unconventional superconducting order, particularly in lower-dimensional systems. Examples are $d$-wave pairing in quasi-2D cuprate superconductors~\cite{Tsuei2000}, $s_{+-}$ pairing in layered iron-pnictides~\cite{Mazin2008,Mazin2009} and $p$-wave triplet pairing in Sr$_2$RuO$_4$~\cite{Mackenzie2003}. Sr$_2$RuO$_4$ is particularly interesting as superconductivity is characterized by a {\it chiral} order parameter that spontaneously breaks time-revisal symmetry~\cite{Nelson2004}, a property it shares with just a few other very low temperature SCs, e.g. UPt$_3$~\cite{Steward1984} and (TMTSF)$_2$PF$_6$~\cite{Jerome2012}. 
Ordering which breaks time reversal has also been discussed in the context of cuprates\citep{Laughlin1998} and Na$_x$CoO$_2$.yH$_2$O\cite{Baskaran2003,Baskaran2004}.
Vortices in these chiral SCs harbor Majorana fermions~\cite{Ivanov2001} which may constitute the building blocks needed for future topological quantum computing technologies, robust against decoherence~\cite{Nayak2008}.

\begin{figure}
\centering
\includegraphics[width=\columnwidth]{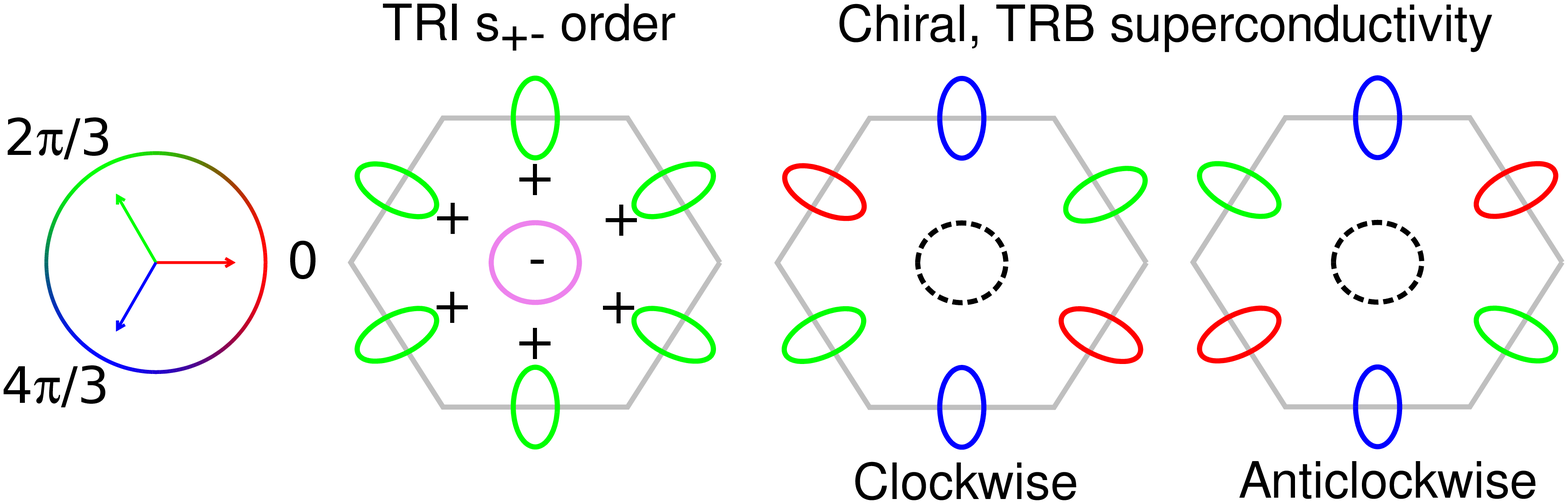}
\caption{Unconventional superconducting orders in a single trilayer of TiSe$_2$. From left to right: colour map representing superconducting phase, Time Reversal Invariant (TRI) s$_{+-}$ ordering, and clockwise and anticlockwise variants of chiral, time reversal broken (TRB) ordering.
}
\label{fig.SCorders}
\end{figure}

Here we focus on the frustrated superconductivity in TiSe$_2$, which in bulk form is a layered semi-metal with a CDW transition at $\sim$200K~\cite{DiSalvo1976}. Upon intercalation with copper, the CDW melts and SC appears with a critical temperature $T_c \approx 4 $K~\cite{Li2007}. In the bulk material, the superconducting order parameter is nodeless~\cite{Li2007}. Using a renormalization group approach, we focus on the case of a single trilayer of TiSe$_2$ and show that it has exciting ordering phenomena. In this case, melting of the CDW phase gives way to one of two possible superconducting ground states, both of which are unconventional. The first is a time-reversal invariant (TRI) state with $s_{+-}$ pairing while the other corresponds to time-reversal broken (TRB), chiral SC, see Fig.~\ref{fig.SCorders}. Their relative stability depends on the precise strength and screening of the electron-electron interactions in 2D trilayer of TiSe$_2$ on top of its substrate.

\paragraph{Effective Lagrangian and couplings} 
A TiSe$_2$ trilayer has an elegant band structure. We have performed {\it ab initio} calculations using FPLO\cite{FPLO_PRB1999,FPLO_url} to find the fermi surfaces. In line with previous reports\cite{Fang1997}, we find (i) two hole-like pockets around the $\Gamma$ point which are nearly degenerate, and (ii) three electron-like pockets, one around each M point in the Brillouin zone. In the 3D case, these bands become elongated along the Z axis and form distorted cylinders -- the 3D material has an additional spherical pocket around the $\Gamma$ point.

Due to approximate nesting between electron and hole bands, there are logarithmic singularities in both particle-particle and particle-hole channels. In order to treat these on an equal footing, we use renormalization group (RG) analysis to establish the low energy couplings. In previously studied cases with nesting and on-site repulsion such as cuprates, pnictides and graphene, RG flow gives low energy couplings that are conducive to SDW order~\cite{Dzyaloshinskii1997,Chubukov2008, Nandkishore2012}. Here however, CDW order arises although the microscopic interactions are repulsive. We will show that this comes about via a special umklapp-process that is allowed by the hexagonal symmetry of TiSe$_2$ which strongly renormalizes the particle-hole and particle-particle channels.

\begin{figure}
\centering
\includegraphics[width=\linewidth]{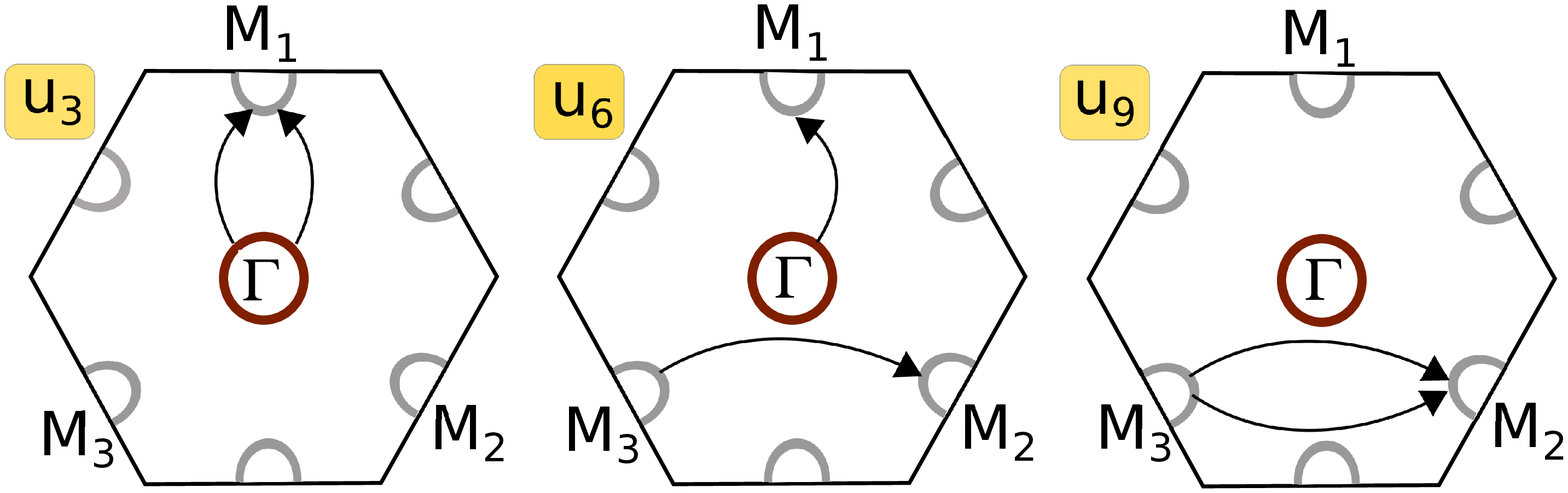}
\caption{Representative umklapp scatterings allowed by the geometry of TiSe$_2$. They arise from the symmetry properties of the M points, viz., $\vec{M}_1 + \vec{M}_2 + \vec{M}_3 =0$ and $2\vec{M}_i \equiv 0$
}
\label{fig.umklapp}
\end{figure}

In the following RG analysis, we approximate the band structure as follows. We merge the two hole pockets around the $\Gamma$-point and give it the band index $0$. With the three electron pockets around the $M$-points, we associate the indices $\alpha=1,2,3$. The electron and hole pockets are approximately nested, so that 
there are nine different scattering processes allowed by momentum conservation (see Supplementary Material for a diagrammatic representation). As the Fermi surfaces have small radii, these couplings can be taken as independent of the precise initial and final momenta. The system is described by the Lagrangian:

\begin{widetext}
\ba
 \mathcal{L} &=& \psi^\dagger_{0}(\partial_\tau - \eps_{0 k}) \psi_{0}+\sum_{\alpha =0}^{3}  \psi^\dagger_{\alpha }(\partial_\tau - \epsilon_{\alpha k}) \psi_{\alpha  }
- \Big\{     U_6 ( \psi^\dagger_{0  } \psi^\dagger_{1  }\psi_{2  } \psi_{3} + \psi^\dagger_{0} \psi^\dagger_{1}\psi_{3 } \psi_{2} + \mathrm{cyclic \phantom{a}exchange} ) \\
\nonumber
&+& \frac{1}{2} U_4 \psi^\dagger_{0  } \psi^\dagger_{0  }\psi_{0  } \psi_{0  } +  \sum_{\alpha =1}^3\big[
    U_1  \psi^\dagger_{0} \psi^\dagger_{\alpha}\psi_{\alpha} \psi_{0}
   + \,     U_2  \psi^\dagger_{0  } \psi^\dagger_{\alpha  }\psi_{0  } \psi_{\alpha  } 
   + \frac{1}{2} U_3  (\psi^\dagger_{0  } \psi^\dagger_{0  }\psi_{\alpha  } \psi_{\alpha  }  + h.c.)
   + \frac{1}{2}U_5 \psi^\dagger_{\alpha  } \psi^\dagger_{\alpha  }\psi_{\alpha  } \psi_{\alpha  } \big]
  \\  \nonumber &+&  \frac{1}{2}  \sum_{\alpha \neq \beta} \big[U_7 \psi^\dagger_{\alpha  } \psi^\dagger_{\beta  }\psi_{\beta  } \psi_{\alpha  }
  +   U_8 \psi^\dagger_{\alpha  } \psi^\dagger_{\beta  }\psi_{\alpha  } \psi_{\beta }
 +  U_9 \psi^\dagger_{\alpha  } \psi^\dagger_{\alpha  }\psi_{\beta  } \psi_{\beta  } \big] \Big\}
\ea
\label{eq.Lagrangian}
\end{widetext}
We have implicitly assumed the spin structure $\sigma \sigma' \sigma' \sigma$ , {\it i.e.} :  $ U_2  \psi^\dagger_{0  } \psi^\dagger_{\alpha  }\psi_{0  } \psi_{\alpha  }  = \sum_{\sigma \sigma'} U_2  \psi^\dagger_{\sigma 0  } \psi^\dagger_{\sigma' \alpha  }\psi_{\sigma' 0  } \psi_{\sigma \alpha  }$. For nested hole and electron pockets the dispersions reduce to
 $(-)\epsilon_{0\bk}  \approx \epsilon_{1\mb{k+M_1}}  = (k_x^2+k_y^2 )/2m - \mu$.   The interactions  $U_3$, $U_6$ and $U_9$ are allowed umklapp processes depicted in Fig.~\ref{fig.umklapp}.
We emphasize that $U_6$  has no analogue in other multi-band systems considered within RG recently, neither in pnictides \cite{Chubukov2008,Thomale2011} nor in graphene \cite{Nandkishore2012,Kiesel2012}. It is allowed by the hexagonal band structure, as the three M momenta add to zero. We later show that precisely this process drives CDW order in TiSe$_2$ as opposed to SDW order in the pnictides or in graphene.

RG flow proceeds by integrating out excitations above a floating cutoff scale. Due to approximate nesting, the electron-hole polarization bubble ( $\vert\Pi_{el-h} \vert \propto \frac{N}{2} \log(\Lambda/\max\{T, \mu_d \} )$) has the same logarithmic divergence as particle-particle bubble ($C_{h-h} =C_{el-el} \propto \frac{N}{2} \log(\Lambda/T )$).
Treating both on an equal footing, we use conventional one-loop RG approach keeping only parquet diagrams.
The flow of couplings is given by:
\bea
\nn \dot{u}_1 &=& u_1^2 + u_3^2 - 2 u_6^2,\\
\nn \dot{u}_2 &=& -2 u_2^2 - 2 u_6^2 + 2 u_2 u_1,\\
\nn \dot{u}_3 &=& u_3 \{ 4 u_1  - 2 u_2 - u_4  - u_5 - 2 u_9\},\\
\nn \dot{u}_4 &=& -u_4^2 - 3 u_3^2, \\
\nn \dot{u}_5 &=& -u_3^2 - u_5^2 - 2 u_9^2, \\
\nn \dot{u}_6 &=& u_6 \{ 2 u_1 - u_2 + u_3 - u_7 - u_8 \}, \\
\nn \dot{u}_7 &=& 2 u_6^2 - u_7^2 - u_8^2, \\
\nn \dot{u}_8 &=& - 2 u_7 u_8, \\
\label{eq.floweqs}\dot{u}_9 &=& -u_3^2 + 2 u_6^2 - 2 u_5u_9 - u_9^2.
\eea
The derivative is with respect to RG time $t=log(W/E)$, where $W$ is the bandwidth and $E$ is the floating RG scale. In addition, we have scaled the interaction amplitudes $U_i$ by the DOS at the Fermi level $N_i$ ($u_i \equiv N_i U_i$). 
The derivation for $u_5$ is illustrated in the Supplementary Material; others can be derived similarly. 
These parquet equations are valid for the energy $E\gtrsim \mu$. Below this energy, density wave channels and superconductivity decouple and the flow has to be modified\cite{Maiti2010}.  
In this letter, we consider small Fermi pockets, thus neglecting the change of flow at $E \sim \mu$.

\paragraph{The leading instabilities}
To investigate the leading instabilities we introduce infinitesimal test vertices in the particle-hole and particle-particle channels:
\ba
&&\delta {\cal{L}}_{CDW}  = \sum_{\alpha=1}^{3} \rho_{c \alpha}^{(0)} \sigma^{0}_{\eta \eta'} \psi^{\dagger}_{0 \eta} \psi_{\alpha \eta'}, \nn  \\
&&\delta {\cal{L}}_{SDW}  = \sum_{\alpha=1}^{3} \rho_{s\alpha}^{(0)} \sigma^x_{\eta \eta'} \psi^{\dagger}_{0 \eta} \psi_{\alpha \eta'},\nn\\
&&\delta {\cal{L}}_{SC}   =   \Delta_{0}^{(0)} i \sigma^y_{\eta \eta'} \psi^{\dagger}_{0 \eta} \psi^\dagger_{0 \eta'} + \sum_{\alpha=1}^{3}  \Delta_{\alpha}^{(0)}i \sigma^y_{\eta \eta'} \psi^{\dagger}_{\alpha \eta} \psi^\dagger_{\alpha \eta'}\nn ,
\ea
where $\sigma^{0}$ and $\sigma^{\alpha}$ are the identity and the Pauli matrices respectively.  We suppose implicit summation over the spin index.
Writing the gap equation for each order, we identify a corresponding `effective vertex' as a function of $u_\alpha$ couplings (see Supplementary Material).
Within this analysis in the framework of the linear approximation, the CDW and SDW orders at each M point decouple. Furthermore, at each M point, 
both CDW and SDW order parameters decouple into two parts which we designate `real' and `imaginary'. They obey correspondingly 
$(\rho_{c/s,\alpha}^{r})^* = +\rho_{c/s,\alpha}^{r} $ and $(\rho_{c/s,\alpha}^{i})^* = -\rho_{c/s,\alpha}^{i}$. 
The effective vertices for real and imaginary SDW and CDW orders are given by
$ \Gamma^{SDW}_{real/imag} = u_1 \pm u_3, $
$ \Gamma^{CDW}_{real/imag} = u_1 \mp u_3 - 2 u_2$.
At lower temperatures, multiple-Q ordering may appear due to interaction between modes. Indeed, such 3Q-ordering has been observed in 3D TiSe$_2$\cite{Ishioka2010}.

In the superconducting channel, our Fermi surface geometry couples the order parameters on individual pockets. 
In accord with symmetry considerations, we get four eigenmodes of superconductivity:
(i)$s_{++}$ conventional superconductivity, characterized by real order parameters on the central pocket ($\Delta_0=\Delta_\Gamma$) and the pockets around M-points ($\Delta_1 =\Delta_2 = \Delta_3=\Delta_M$), both having the same sign ($sign(\Delta_\Gamma) =sign(\Delta_M)$). (ii) $s_{+-}$ with real order parameters having different signs on the central and M-pockets, i.e., ($sign(\Delta_0) = -sign(\Delta_M)$), as shown in Fig.~\ref{fig.SCorders}. It is analogous to the order parameter proposed for the recently discovered Fe-based superconductors.
(iii \& iv) chiral superconductivity, which breaks time reversal symmetry.
At the level of linearized gap equations, the central pocket is completely decoupled. 
There are two degenerate solutions, corresponding to clockwise and anticlockwise winding of the phase of the order parameters, shown in Fig.~\ref{fig.SCorders}.
One of the two solutions is given by $\Delta_1=e^{i2\pi/3} \Delta_2 = e^{-i2\pi/3} \Delta_3 = \Delta_M$. 
A similar phase has been proposed in highly doped graphene\cite{Doniach2007,Pathak2010,Nandkishore2012,Kiesel2012}. 
The effective vertices are given by
$ -\Gamma_{s_{++}}^{SC} = -(u_4 + u_5 + 2u_9)/2 -sign(u_3) R  $,
$ -\Gamma_{s_{+-}}^{SC} = -(u_4 + u_5 + 2u_9)/2 +sign(u_3)  R $,
$-\Gamma_{chiral}^{SC}  = -u_5 + u_9$,
where we have denoted $R = \sqrt{12 u_3^2 + (u_4-u_5-2u_9)^2}/2 $. 

\begin{figure}
\centering
\includegraphics[width=\columnwidth]{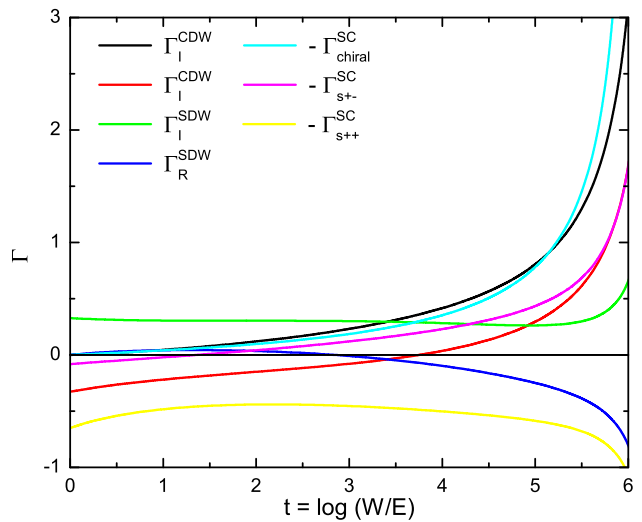}
\caption{RG flow of effective vertices. We have used bare interactions estimated assuming intra-atomic Coulomb interactions: $u^{(0)}_1 = u^{(0)}_2 = u^{(0)}_3=0.82u^{(0)} $, $ u^{(0)}_4 = 1.85u^{(0)}$, $u^{(0)}_5 =u^{(0)}_7=u^{(0)}_8=u^{(0)}_9= 0.6u^{(0)}$ and $ u^{(0)}_6 = 0.675u^{(0)}$, taking $u^{(0)}=0.2$ . Chiral SC eventually dominates.}
\label{fig.RGflow}
\end{figure}

We first analyze the behaviour of the system by treating the `bare interactions' in mean field. As the interactions are dominated by intra-atomic Coulomb repulsion the bare couplings  are proportional to the partial contribution of Ti t$_{2g}$ orbitals and Se $p$ orbitals to DOS in the Fermi pockets. 
From \textit{ab initio }calculations, we find the orbital contributions to states in each Fermi pocket to be $N^{\Gamma}_{Ti}\sim 0.8$, $N^{\Gamma}_{Se}\sim 1.1$, $N^{M}_{Ti}\sim 0.75$
and $N^{M}_{Se}\sim 0.2$. We can now estimate the bare interactions, e.g., 
$u_1^{(0)} = u^{(0)} ( N^\Gamma_{Ti} N^M_{Ti}+ N^\Gamma_{Se}
 N^M_{Se})$,  $u_4^{(0)} = u^{(0)} ( \{N^\Gamma_{Ti}\}^2 + 
 \{N^\Gamma_{Se}\}^2)$, where $u^{(0)}$ is a parameter capturing the strength of the Coulomb interaction. Using these values, we find that the largest effective vertex corresponds to real SDW order 
$\Gamma^{SDW}_{real} \sim (0.33)u^{(0)} $. Superconducting channels drop out as their effective vertices are repulsive.
Mean field treatment thus predicts SDW order; however, RG flow modifies the couplings and changes the preferred ordering. Fig.~\ref{fig.RGflow} shows the RG flow of effective vertices starting from these bare interactions -- chiral SC ultimately dominates. 

\begin{figure}
\centering
\includegraphics[width=0.9\columnwidth]{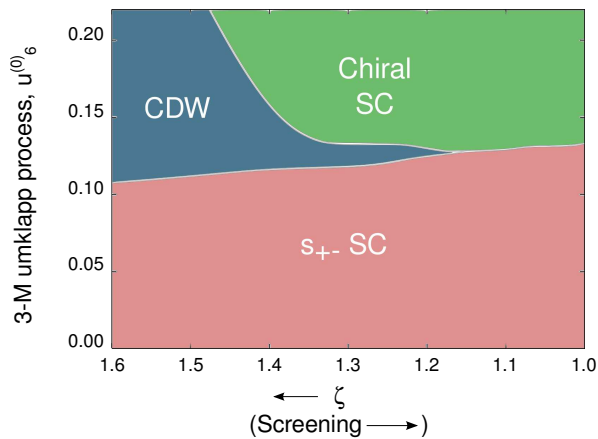}
\caption{The basins of the three fixed points. The dashed line depicts $u_6^{bare} (\zeta)$. The bare vertices are $u^{(0)}_1/\zeta = u^{(0)}_2 = u^{(0)}_3=0.82u^{(0)}/\zeta $, $ u^{(0)}_4 = 1.85u^{(0)}$, $u^{(0)}_5/\zeta =u^{(0)}_7/\zeta=u^{(0)}_8=u^{(0)}_9= 0.6u^{(0)}/\zeta$. The precise location of basin boundaries weakly depends on $u^{(0)}$.}
\label{fig.phasediagram}
\end{figure}

\paragraph{Fixed points in the RG flow}
The flow of couplings given by Eqs.~\ref{eq.floweqs} is governed by three fixed points, wherein all couplings scale with with one diverging quantity. We rewrite the interactions as $u_i = b_i u$ with $u>0$ being the divergent scale. The three fixed points are:

(i)``CDW fixed point": $b_2 =-1$, with all other couplings negligible $b_i =0$. At this fixed point, the largest effective vertices correspond to both real and imaginary solutions of CDW order $\Gamma^{CDW}_{real}=\Gamma^{CDW}_{imag}$.

(ii) ``Chiral SC fixed point":  $b_9 = -b_5/2>0$, while other $b$'s vanish. The largest effective vertex then corresponds to chiral SC. 

(iii) ``$s_{+-}$ fixed point'': 
$ b_4 = -1$,
$ b_3 = \sqrt{5/11} $,
$ b_1 =  1/11$,
$ b_5,b_9 = -1/3$.
The couplings $b_2$, $b_6$, $b_7$, $b_8$ vanish. The leading vertex is $s_{+-}$ SC.

If ordering in TiSe$_2$ were driven by phonons, we would expect to see CDW order and $s_{++}$ superconductivity. In contrast, electronic correlations under RG flow give CDW order and $s_{+-}$ or chiral superconductivity. 
The absence of $s_{++}$ superconductivity can be traced to the flow equation for $u_3$ in Eqs.~\ref{eq.floweqs}. The sign of $u_3$ cannot change under RG flow and always remains positive, thus favouring $s_{+-}$ pairing over $s_{++}$. 

\paragraph{Phase diagram in RG scheme}
RG flow depends on the initial, bare couplings which we estimate using \textit{ab initio }data for the orbital DOS. Building upon this, we introduce two free parameters, $u_6^0$ and $\zeta$, to characterize the bare couplings. 
The parameter $u_6^0$ is simply the bare value of the $u_6$ coupling; we use it as a parameter in order to emphasize the key role of the $u_6$ process. The second parameter $\zeta$ models the momentum dependence of the screened Coulomb interaction. 
The low energy scattering processes fall into two classes: small and large ($\sim M$) momentum transfer. The latter are reduced by the factor $\zeta$. For example, we have  $u_1^{(0)} = u^{(0)} ( N^\Gamma_{Ti} N^M_{Ti}+ N^\Gamma_{Se} N^M_{Se})$ and $u_2^{(0)} = u^{(0)} ( N^\Gamma_{Ti} N^M_{Ti}+ N^\Gamma_{Se} N^M_{Se})/\zeta$. For strong screening, we expect local interactions and momentum-independent interactions, giving $\zeta\sim 1$. For weak screening, $\zeta > 1$.

Fig.~\ref{fig.phasediagram} shows the fate of RG flow as a function of these two parameters. The crucial role of hexagonal symmetry can be seen by examining the line $u_6^0 = 0$. When the bare value of $u_6$ is zero, RG flow cannot generate a finite $u_6$ value (see Eqs.~\ref{eq.floweqs}). Without $u_6$ (along the $u_6^0 =0$ line), we do not approach the CDW fixed point or the chiral SC fixed point.
To estimate the `microscopic' value of $u_6^0$, we use the same reasoning as with the other bare parameters to obtain 
$
u_6^{bare}(\zeta) = \{ (N^\Gamma_{Se})^{1/2}  (N^M_{Se})^{3/2} + (N^\Gamma_{Ti})^{1/2} (N^M_{Ti})^{3/2} \}/\zeta
$. As $u_6$ involves large momentum transfer, it is scaled down by $\zeta$.
This choice of $u_6$ places us in the basins of CDW and $s_{+-}$ fixed points. However, for some $\zeta$ values, the microscopic parameters lie very close to the border of the CDW basin.
Taken together, our results suggest that the ground state of 2D TiSe$_2$ could have CDW order, chiral superconductivity or $s_{+-}$ pairing. 

\paragraph{Discussion}
Our results for TiSe$_2$ should be compared with the pnictides wherein Coulomb interactions lead to $s_{+-}$ SC order which competes with SDW order. Ultimately, this difference in behaviour stems from $u_6$, the umklapp process allowed by the geometry of the M points. CDW order requires a negative value of $u_2$ at the fixed point, whereas the bare Coulombic value of $u_2$ is positive.  However, a non-zero $u_6$ reduces the value of $u_2$ under RG flow (see Eq.~\ref{eq.floweqs}) to negative values.

The chiral superconducting state breaks time reversal symmetry with many interesting consequences. Previous studies have highlighted the presence of fractional vortices and domain walls in such a superconducting state~\cite{Yanagisawa2012}. If the central pocket is indeed decoupled as shown in Fig.~\ref{fig.SCorders}, this pocket may undergo pairing with a different transition temperature. Instead, the central pocket could possess $d+id$ order! This possibility is also favoured by \textit{ab initio } calculations which show that each M pocket is dominated by a single Ti t$_{2g}$ orbital. The central pocket shows a strong angular dependence in $t_{2g}$ orbital character consistent with $d+id$ pairing.

\paragraph{Summary}
We have analyzed the competing phases in two dimensional hexagonal structures which allow special umklapp processes. In systems with Coulomb repulsion, these processes give rise to CDW order instead of SDW. This CDW state competes with chiral and $s_{+-}$ superconductivity and \textit{not} $s_{++}$ superconductivity expected from a phonon mechanism.  
We have focused on two dimensional TiSe$_2$ with two small hole pockets around $\Gamma$ point and electron pockets around the M-points, but it is interesting to note that 2D  TiS$_2$ actually has a similar band structure. Layered 3D TiSe$_2$ has an additional spherical hole pocket around the $\Gamma$ point -- our RG group analysis is still valid, but for high energies when the bands are 2D-like. 
While nodeless superconductivity has been seen in 3D TiSe$_2$\cite{Morosan2006,Li2007}, our results call for a more detailed examination of the nature of superconductivity, in particular in exfoliated layers of this materials with nanoscopic thicknesses.


\begin{thebibliography}{33}
\expandafter\ifx\csname natexlab\endcsname\relax\def\natexlab#1{#1}\fi
\expandafter\ifx\csname bibnamefont\endcsname\relax
  \def\bibnamefont#1{#1}\fi
\expandafter\ifx\csname bibfnamefont\endcsname\relax
  \def\bibfnamefont#1{#1}\fi
\expandafter\ifx\csname citenamefont\endcsname\relax
  \def\citenamefont#1{#1}\fi
\expandafter\ifx\csname url\endcsname\relax
  \def\url#1{\texttt{#1}}\fi
\expandafter\ifx\csname urlprefix\endcsname\relax\def\urlprefix{URL }\fi
\providecommand{\bibinfo}[2]{#2}
\providecommand{\eprint}[2][]{\url{#2}}

\bibitem[{\citenamefont{Xu et~al.}(2013)\citenamefont{Xu, Liang, Shi, and
  Chen}}]{Xu2013}
\bibinfo{author}{\bibfnamefont{M.}~\bibnamefont{Xu}},
  \bibinfo{author}{\bibfnamefont{T.}~\bibnamefont{Liang}},
  \bibinfo{author}{\bibfnamefont{M.}~\bibnamefont{Shi}}, \bibnamefont{and}
  \bibinfo{author}{\bibfnamefont{H.}~\bibnamefont{Chen}},
  \bibinfo{journal}{Chemical Reviews} \textbf{\bibinfo{volume}{113}},
  \bibinfo{pages}{3766} (\bibinfo{year}{2013}),
  \eprint{http://pubs.acs.org/doi/pdf/10.1021/cr300263a},
  \urlprefix\url{http://pubs.acs.org/doi/abs/10.1021/cr300263a}.

\bibitem[{\citenamefont{Radisavljevic and Kis}(2013)}]{Radisavljevic2013}
\bibinfo{author}{\bibfnamefont{B.}~\bibnamefont{Radisavljevic}}
  \bibnamefont{and} \bibinfo{author}{\bibfnamefont{A.}~\bibnamefont{Kis}},
  \bibinfo{journal}{Nat Mater} \textbf{\bibinfo{volume}{12}},
  \bibinfo{pages}{815} (\bibinfo{year}{2013}), ISSN \bibinfo{issn}{1476-1122},
  \urlprefix\url{http://dx.doi.org/10.1038/nmat3687}.

\bibitem[{\citenamefont{Humberto~Terrones}(2013)}]{HumbertoTerrones2013}
\bibinfo{author}{\bibfnamefont{F.~L. p.-U. M.~T.}
  \bibnamefont{Humberto~Terrones}}, \bibinfo{journal}{Nature: Scientific
  Reports} \textbf{\bibinfo{volume}{3}}, \bibinfo{pages}{1549}
  (\bibinfo{year}{2013}).

\bibitem[{\citenamefont{Wang et~al.}(2012)\citenamefont{Wang, Kalantar-Zadeh,
  Kis, Coleman, and Strano}}]{wang2012}
\bibinfo{author}{\bibfnamefont{Q.~H.} \bibnamefont{Wang}},
  \bibinfo{author}{\bibfnamefont{K.}~\bibnamefont{Kalantar-Zadeh}},
  \bibinfo{author}{\bibfnamefont{A.}~\bibnamefont{Kis}},
  \bibinfo{author}{\bibfnamefont{J.~N.} \bibnamefont{Coleman}},
  \bibnamefont{and} \bibinfo{author}{\bibfnamefont{M.~S.}
  \bibnamefont{Strano}}, \bibinfo{journal}{Nat Nano}
  \textbf{\bibinfo{volume}{7}}, \bibinfo{pages}{699} (\bibinfo{year}{2012}),
  ISSN \bibinfo{issn}{1748-3387},
  \urlprefix\url{http://dx.doi.org/10.1038/nnano.2012.193}.

\bibitem[{\citenamefont{Coleman et~al.}(2011)\citenamefont{Coleman, Lotya,
  O'Neill, Bergin, King, Khan, Young, Gaucher, De, Smith et~al.}}]{Coleman2011}
\bibinfo{author}{\bibfnamefont{J.~N.} \bibnamefont{Coleman}},
  \bibinfo{author}{\bibfnamefont{M.}~\bibnamefont{Lotya}},
  \bibinfo{author}{\bibfnamefont{A.}~\bibnamefont{O'Neill}},
  \bibinfo{author}{\bibfnamefont{S.~D.} \bibnamefont{Bergin}},
  \bibinfo{author}{\bibfnamefont{P.~J.} \bibnamefont{King}},
  \bibinfo{author}{\bibfnamefont{U.}~\bibnamefont{Khan}},
  \bibinfo{author}{\bibfnamefont{K.}~\bibnamefont{Young}},
  \bibinfo{author}{\bibfnamefont{A.}~\bibnamefont{Gaucher}},
  \bibinfo{author}{\bibfnamefont{S.}~\bibnamefont{De}},
  \bibinfo{author}{\bibfnamefont{R.~J.} \bibnamefont{Smith}},
  \bibnamefont{et~al.}, \bibinfo{journal}{Science}
  \textbf{\bibinfo{volume}{331}}, \bibinfo{pages}{568} (\bibinfo{year}{2011}),
  \eprint{http://www.sciencemag.org/content/331/6017/568.full.pdf},
  \urlprefix\url{http://www.sciencemag.org/content/331/6017/568.abstract}.

\bibitem[{\citenamefont{Tsuei and Kirtley}(2000)}]{Tsuei2000}
\bibinfo{author}{\bibfnamefont{C.~C.} \bibnamefont{Tsuei}} \bibnamefont{and}
  \bibinfo{author}{\bibfnamefont{J.~R.} \bibnamefont{Kirtley}},
  \bibinfo{journal}{Rev. Mod. Phys.} \textbf{\bibinfo{volume}{72}},
  \bibinfo{pages}{969} (\bibinfo{year}{2000}),
  \urlprefix\url{http://link.aps.org/doi/10.1103/RevModPhys.72.969}.

\bibitem[{\citenamefont{Mazin et~al.}(2008)\citenamefont{Mazin, Singh,
  Johannes, and Du}}]{Mazin2008}
\bibinfo{author}{\bibfnamefont{I.~I.} \bibnamefont{Mazin}},
  \bibinfo{author}{\bibfnamefont{D.~J.} \bibnamefont{Singh}},
  \bibinfo{author}{\bibfnamefont{M.~D.} \bibnamefont{Johannes}},
  \bibnamefont{and} \bibinfo{author}{\bibfnamefont{M.~H.} \bibnamefont{Du}},
  \bibinfo{journal}{Phys. Rev. Lett.} \textbf{\bibinfo{volume}{101}},
  \bibinfo{pages}{057003} (\bibinfo{year}{2008}),
  \urlprefix\url{http://link.aps.org/doi/10.1103/PhysRevLett.101.057003}.

\bibitem[{\citenamefont{Mazin and Schmalian}(2009)}]{Mazin2009}
\bibinfo{author}{\bibfnamefont{I.}~\bibnamefont{Mazin}} \bibnamefont{and}
  \bibinfo{author}{\bibfnamefont{J.}~\bibnamefont{Schmalian}},
  \bibinfo{journal}{Physica C: Superconductivity}
  \textbf{\bibinfo{volume}{469}}, \bibinfo{pages}{614 } (\bibinfo{year}{2009}),
  ISSN \bibinfo{issn}{0921-4534}, \bibinfo{note}{superconductivity in
  Iron-Pnictides},
  \urlprefix\url{http://www.sciencedirect.com/science/article/pii/S09214534090%
01002}.

\bibitem[{\citenamefont{Mackenzie and Maeno}(2003)}]{Mackenzie2003}
\bibinfo{author}{\bibfnamefont{A.~P.} \bibnamefont{Mackenzie}}
  \bibnamefont{and} \bibinfo{author}{\bibfnamefont{Y.}~\bibnamefont{Maeno}},
  \bibinfo{journal}{Rev. Mod. Phys.} \textbf{\bibinfo{volume}{75}},
  \bibinfo{pages}{657} (\bibinfo{year}{2003}),
  \urlprefix\url{http://link.aps.org/doi/10.1103/RevModPhys.75.657}.

\bibitem[{\citenamefont{Nelson et~al.}(2004)\citenamefont{Nelson, Mao, Maeno,
  and Liu}}]{Nelson2004}
\bibinfo{author}{\bibfnamefont{K.~D.} \bibnamefont{Nelson}},
  \bibinfo{author}{\bibfnamefont{Z.~Q.} \bibnamefont{Mao}},
  \bibinfo{author}{\bibfnamefont{Y.}~\bibnamefont{Maeno}}, \bibnamefont{and}
  \bibinfo{author}{\bibfnamefont{Y.}~\bibnamefont{Liu}},
  \bibinfo{journal}{Science} \textbf{\bibinfo{volume}{306}},
  \bibinfo{pages}{1151} (\bibinfo{year}{2004}),
  \eprint{http://www.sciencemag.org/content/306/5699/1151.full.pdf},
  \urlprefix\url{http://www.sciencemag.org/content/306/5699/1151.abstract}.

\bibitem[{\citenamefont{Stewart et~al.}(1984)\citenamefont{Stewart, Fisk,
  Willis, and Smith}}]{Steward1984}
\bibinfo{author}{\bibfnamefont{G.~R.} \bibnamefont{Stewart}},
  \bibinfo{author}{\bibfnamefont{Z.}~\bibnamefont{Fisk}},
  \bibinfo{author}{\bibfnamefont{J.~O.} \bibnamefont{Willis}},
  \bibnamefont{and} \bibinfo{author}{\bibfnamefont{J.~L.} \bibnamefont{Smith}},
  \bibinfo{journal}{Phys. Rev. Lett.} \textbf{\bibinfo{volume}{52}},
  \bibinfo{pages}{679} (\bibinfo{year}{1984}),
  \urlprefix\url{http://link.aps.org/doi/10.1103/PhysRevLett.52.679}.

\bibitem[{\citenamefont{{J\'erome, D.} et~al.}(1980)\citenamefont{{J\'erome,
  D.}, {Mazaud, A.}, {Ribault, M.}, and {Bechgaard, K.}}}]{Jerome2012}
\bibinfo{author}{\bibnamefont{{J\'erome, D.}}},
  \bibinfo{author}{\bibnamefont{{Mazaud, A.}}},
  \bibinfo{author}{\bibnamefont{{Ribault, M.}}}, \bibnamefont{and}
  \bibinfo{author}{\bibnamefont{{Bechgaard, K.}}}, \bibinfo{journal}{J.
  Physique Lett.} \textbf{\bibinfo{volume}{41}}, \bibinfo{pages}{95}
  (\bibinfo{year}{1980}),
  \urlprefix\url{http://dx.doi.org/10.1051/jphyslet:0198000410409500}.

\bibitem[{\citenamefont{Laughlin}(1998)}]{Laughlin1998}
\bibinfo{author}{\bibfnamefont{R.~B.} \bibnamefont{Laughlin}},
  \bibinfo{journal}{Phys. Rev. Lett.} \textbf{\bibinfo{volume}{80}},
  \bibinfo{pages}{5188} (\bibinfo{year}{1998}),
  \urlprefix\url{http://link.aps.org/doi/10.1103/PhysRevLett.80.5188}.

\bibitem[{\citenamefont{Baskaran}(2003)}]{Baskaran2003}
\bibinfo{author}{\bibfnamefont{G.}~\bibnamefont{Baskaran}},
  \bibinfo{journal}{Phys. Rev. Lett.} \textbf{\bibinfo{volume}{91}},
  \bibinfo{pages}{097003} (\bibinfo{year}{2003}),
  \urlprefix\url{http://link.aps.org/doi/10.1103/PhysRevLett.91.097003}.

\bibitem[{\citenamefont{Sa et~al.}(2004)\citenamefont{Sa, Sardar, and
  Baskaran}}]{Baskaran2004}
\bibinfo{author}{\bibfnamefont{D.}~\bibnamefont{Sa}},
  \bibinfo{author}{\bibfnamefont{M.}~\bibnamefont{Sardar}}, \bibnamefont{and}
  \bibinfo{author}{\bibfnamefont{G.}~\bibnamefont{Baskaran}},
  \bibinfo{journal}{Phys. Rev. B} \textbf{\bibinfo{volume}{70}},
  \bibinfo{pages}{104505} (\bibinfo{year}{2004}),
  \urlprefix\url{http://link.aps.org/doi/10.1103/PhysRevB.70.104505}.

\bibitem[{\citenamefont{Ivanov}(2001)}]{Ivanov2001}
\bibinfo{author}{\bibfnamefont{D.~A.} \bibnamefont{Ivanov}},
  \bibinfo{journal}{Phys. Rev. Lett.} \textbf{\bibinfo{volume}{86}},
  \bibinfo{pages}{268} (\bibinfo{year}{2001}),
  \urlprefix\url{http://link.aps.org/doi/10.1103/PhysRevLett.86.268}.

\bibitem[{\citenamefont{Nayak et~al.}(2008)\citenamefont{Nayak, Simon, Stern,
  Freedman, and Das~Sarma}}]{Nayak2008}
\bibinfo{author}{\bibfnamefont{C.}~\bibnamefont{Nayak}},
  \bibinfo{author}{\bibfnamefont{S.~H.} \bibnamefont{Simon}},
  \bibinfo{author}{\bibfnamefont{A.}~\bibnamefont{Stern}},
  \bibinfo{author}{\bibfnamefont{M.}~\bibnamefont{Freedman}}, \bibnamefont{and}
  \bibinfo{author}{\bibfnamefont{S.}~\bibnamefont{Das~Sarma}},
  \bibinfo{journal}{Rev. Mod. Phys.} \textbf{\bibinfo{volume}{80}},
  \bibinfo{pages}{1083} (\bibinfo{year}{2008}),
  \urlprefix\url{http://link.aps.org/doi/10.1103/RevModPhys.80.1083}.

\bibitem[{\citenamefont{Di~Salvo et~al.}(1976)\citenamefont{Di~Salvo, Moncton,
  and Waszczak}}]{DiSalvo1976}
\bibinfo{author}{\bibfnamefont{F.~J.} \bibnamefont{Di~Salvo}},
  \bibinfo{author}{\bibfnamefont{D.~E.} \bibnamefont{Moncton}},
  \bibnamefont{and} \bibinfo{author}{\bibfnamefont{J.~V.}
  \bibnamefont{Waszczak}}, \bibinfo{journal}{Phys. Rev. B}
  \textbf{\bibinfo{volume}{14}}, \bibinfo{pages}{4321} (\bibinfo{year}{1976}),
  \urlprefix\url{http://link.aps.org/doi/10.1103/PhysRevB.14.4321}.

\bibitem[{\citenamefont{Li et~al.}(2007)\citenamefont{Li, Wu, Chen, and
  Taillefer}}]{Li2007}
\bibinfo{author}{\bibfnamefont{S.~Y.} \bibnamefont{Li}},
  \bibinfo{author}{\bibfnamefont{G.}~\bibnamefont{Wu}},
  \bibinfo{author}{\bibfnamefont{X.~H.} \bibnamefont{Chen}}, \bibnamefont{and}
  \bibinfo{author}{\bibfnamefont{L.}~\bibnamefont{Taillefer}},
  \bibinfo{journal}{Phys. Rev. Lett.} \textbf{\bibinfo{volume}{99}},
  \bibinfo{pages}{107001} (\bibinfo{year}{2007}),
  \urlprefix\url{http://link.aps.org/doi/10.1103/PhysRevLett.99.107001}.

\bibitem[{\citenamefont{Koepernik and Eschrig}(1999)}]{FPLO_PRB1999}
\bibinfo{author}{\bibfnamefont{K.}~\bibnamefont{Koepernik}} \bibnamefont{and}
  \bibinfo{author}{\bibfnamefont{H.}~\bibnamefont{Eschrig}},
  \bibinfo{journal}{Phys. Rev. B} \textbf{\bibinfo{volume}{59}},
  \bibinfo{pages}{1743} (\bibinfo{year}{1999}),
  \urlprefix\url{http://link.aps.org/doi/10.1103/PhysRevB.59.1743}.

\bibitem[{FPL()}]{FPLO_url}
\urlprefix\url{http://www.fplo.de}.

\bibitem[{\citenamefont{Fang et~al.}(1997)\citenamefont{Fang, de~Groot, and
  Haas}}]{Fang1997}
\bibinfo{author}{\bibfnamefont{C.~M.} \bibnamefont{Fang}},
  \bibinfo{author}{\bibfnamefont{R.~A.} \bibnamefont{de~Groot}},
  \bibnamefont{and} \bibinfo{author}{\bibfnamefont{C.}~\bibnamefont{Haas}},
  \bibinfo{journal}{Phys. Rev. B} \textbf{\bibinfo{volume}{56}},
  \bibinfo{pages}{4455} (\bibinfo{year}{1997}),
  \urlprefix\url{http://link.aps.org/doi/10.1103/PhysRevB.56.4455}.

\bibitem[{\citenamefont{Zheleznyak et~al.}(1997)\citenamefont{Zheleznyak,
  Yakovenko, and Dzyaloshinskii}}]{Dzyaloshinskii1997}
\bibinfo{author}{\bibfnamefont{A.~T.} \bibnamefont{Zheleznyak}},
  \bibinfo{author}{\bibfnamefont{V.~M.} \bibnamefont{Yakovenko}},
  \bibnamefont{and} \bibinfo{author}{\bibfnamefont{I.~E.}
  \bibnamefont{Dzyaloshinskii}}, \bibinfo{journal}{Phys. Rev. B}
  \textbf{\bibinfo{volume}{55}}, \bibinfo{pages}{3200} (\bibinfo{year}{1997}),
  \urlprefix\url{http://link.aps.org/doi/10.1103/PhysRevB.55.3200}.

\bibitem[{\citenamefont{Chubukov et~al.}(2008)\citenamefont{Chubukov, Efremov,
  and Eremin}}]{Chubukov2008}
\bibinfo{author}{\bibfnamefont{A.~V.} \bibnamefont{Chubukov}},
  \bibinfo{author}{\bibfnamefont{D.~V.} \bibnamefont{Efremov}},
  \bibnamefont{and} \bibinfo{author}{\bibfnamefont{I.}~\bibnamefont{Eremin}},
  \bibinfo{journal}{Phys. Rev. B} \textbf{\bibinfo{volume}{78}},
  \bibinfo{pages}{134512} (\bibinfo{year}{2008}),
  \urlprefix\url{http://link.aps.org/doi/10.1103/PhysRevB.78.134512}.

\bibitem[{\citenamefont{{Nandkishore} et~al.}(2012)\citenamefont{{Nandkishore},
  {Levitov}, and {Chubukov}}}]{Nandkishore2012}
\bibinfo{author}{\bibfnamefont{R.}~\bibnamefont{{Nandkishore}}},
  \bibinfo{author}{\bibfnamefont{L.~S.} \bibnamefont{{Levitov}}},
  \bibnamefont{and} \bibinfo{author}{\bibfnamefont{A.~V.}
  \bibnamefont{{Chubukov}}}, \bibinfo{journal}{Nature Physics}
  \textbf{\bibinfo{volume}{8}}, \bibinfo{pages}{158} (\bibinfo{year}{2012}),
  \eprint{1107.1903}.

\bibitem[{\citenamefont{Thomale et~al.}(2011)\citenamefont{Thomale, Platt,
  Hanke, and Bernevig}}]{Thomale2011}
\bibinfo{author}{\bibfnamefont{R.}~\bibnamefont{Thomale}},
  \bibinfo{author}{\bibfnamefont{C.}~\bibnamefont{Platt}},
  \bibinfo{author}{\bibfnamefont{W.}~\bibnamefont{Hanke}}, \bibnamefont{and}
  \bibinfo{author}{\bibfnamefont{B.~A.} \bibnamefont{Bernevig}},
  \bibinfo{journal}{Phys. Rev. Lett.} \textbf{\bibinfo{volume}{106}},
  \bibinfo{pages}{187003} (\bibinfo{year}{2011}),
  \urlprefix\url{http://link.aps.org/doi/10.1103/PhysRevLett.106.187003}.

\bibitem[{\citenamefont{Kiesel et~al.}(2012)\citenamefont{Kiesel, Platt, Hanke,
  Abanin, and Thomale}}]{Kiesel2012}
\bibinfo{author}{\bibfnamefont{M.~L.} \bibnamefont{Kiesel}},
  \bibinfo{author}{\bibfnamefont{C.}~\bibnamefont{Platt}},
  \bibinfo{author}{\bibfnamefont{W.}~\bibnamefont{Hanke}},
  \bibinfo{author}{\bibfnamefont{D.~A.} \bibnamefont{Abanin}},
  \bibnamefont{and} \bibinfo{author}{\bibfnamefont{R.}~\bibnamefont{Thomale}},
  \bibinfo{journal}{Phys. Rev. B} \textbf{\bibinfo{volume}{86}},
  \bibinfo{pages}{020507} (\bibinfo{year}{2012}),
  \urlprefix\url{http://link.aps.org/doi/10.1103/PhysRevB.86.020507}.

\bibitem[{\citenamefont{Maiti and Chubukov}(2010)}]{Maiti2010}
\bibinfo{author}{\bibfnamefont{S.}~\bibnamefont{Maiti}} \bibnamefont{and}
  \bibinfo{author}{\bibfnamefont{A.~V.} \bibnamefont{Chubukov}},
  \bibinfo{journal}{Phys. Rev. B} \textbf{\bibinfo{volume}{82}},
  \bibinfo{pages}{214515} (\bibinfo{year}{2010}),
  \urlprefix\url{http://link.aps.org/doi/10.1103/PhysRevB.17.1839}.

\bibitem[{\citenamefont{Ishioka et~al.}(2010)\citenamefont{Ishioka, Liu,
  Shimatake, Kurosawa, Ichimura, Toda, Oda, and Tanda}}]{Ishioka2010}
\bibinfo{author}{\bibfnamefont{J.}~\bibnamefont{Ishioka}},
  \bibinfo{author}{\bibfnamefont{Y.~H.} \bibnamefont{Liu}},
  \bibinfo{author}{\bibfnamefont{K.}~\bibnamefont{Shimatake}},
  \bibinfo{author}{\bibfnamefont{T.}~\bibnamefont{Kurosawa}},
  \bibinfo{author}{\bibfnamefont{K.}~\bibnamefont{Ichimura}},
  \bibinfo{author}{\bibfnamefont{Y.}~\bibnamefont{Toda}},
  \bibinfo{author}{\bibfnamefont{M.}~\bibnamefont{Oda}}, \bibnamefont{and}
  \bibinfo{author}{\bibfnamefont{S.}~\bibnamefont{Tanda}},
  \bibinfo{journal}{Phys. Rev. Lett.} \textbf{\bibinfo{volume}{105}},
  \bibinfo{pages}{176401} (\bibinfo{year}{2010}),
  \urlprefix\url{http://link.aps.org/doi/10.1103/PhysRevLett.105.176401}.

\bibitem[{\citenamefont{Black-Schaffer and Doniach}(2007)}]{Doniach2007}
\bibinfo{author}{\bibfnamefont{A.~M.} \bibnamefont{Black-Schaffer}}
  \bibnamefont{and} \bibinfo{author}{\bibfnamefont{S.}~\bibnamefont{Doniach}},
  \bibinfo{journal}{Phys. Rev. B} \textbf{\bibinfo{volume}{75}},
  \bibinfo{pages}{134512} (\bibinfo{year}{2007}),
  \urlprefix\url{http://link.aps.org/doi/10.1103/PhysRevB.75.134512}.

\bibitem[{\citenamefont{Pathak et~al.}(2010)\citenamefont{Pathak, Shenoy, and
  Baskaran}}]{Pathak2010}
\bibinfo{author}{\bibfnamefont{S.}~\bibnamefont{Pathak}},
  \bibinfo{author}{\bibfnamefont{V.~B.} \bibnamefont{Shenoy}},
  \bibnamefont{and} \bibinfo{author}{\bibfnamefont{G.}~\bibnamefont{Baskaran}},
  \bibinfo{journal}{Phys. Rev. B} \textbf{\bibinfo{volume}{81}},
  \bibinfo{pages}{085431} (\bibinfo{year}{2010}),
  \urlprefix\url{http://link.aps.org/doi/10.1103/PhysRevB.81.085431}.

\bibitem[{\citenamefont{Yanagisawa et~al.}(2012)\citenamefont{Yanagisawa,
  Tanaka, Hase, and Yamaji}}]{Yanagisawa2012}
\bibinfo{author}{\bibfnamefont{T.}~\bibnamefont{Yanagisawa}},
  \bibinfo{author}{\bibfnamefont{Y.}~\bibnamefont{Tanaka}},
  \bibinfo{author}{\bibfnamefont{I.}~\bibnamefont{Hase}}, \bibnamefont{and}
  \bibinfo{author}{\bibfnamefont{K.}~\bibnamefont{Yamaji}},
  \bibinfo{journal}{Journal of the Physical Society of Japan}
  \textbf{\bibinfo{volume}{81}}, \bibinfo{pages}{024712}
  (\bibinfo{year}{2012}),
  \urlprefix\url{http://jpsj.ipap.jp/link?JPSJ/81/024712/}.

\bibitem[{\citenamefont{Morosan et~al.}(2006)\citenamefont{Morosan, Zandbergen,
  Dennis, Bos, Onose, Klimczuk, Ramirez, Ong, and Cava}}]{Morosan2006}
\bibinfo{author}{\bibfnamefont{E.}~\bibnamefont{Morosan}},
  \bibinfo{author}{\bibfnamefont{H.~W.} \bibnamefont{Zandbergen}},
  \bibinfo{author}{\bibfnamefont{B.~S.} \bibnamefont{Dennis}},
  \bibinfo{author}{\bibfnamefont{J.~W.~G.} \bibnamefont{Bos}},
  \bibinfo{author}{\bibfnamefont{Y.}~\bibnamefont{Onose}},
  \bibinfo{author}{\bibfnamefont{T.}~\bibnamefont{Klimczuk}},
  \bibinfo{author}{\bibfnamefont{A.~P.} \bibnamefont{Ramirez}},
  \bibinfo{author}{\bibfnamefont{N.~P.} \bibnamefont{Ong}}, \bibnamefont{and}
  \bibinfo{author}{\bibfnamefont{R.~J.} \bibnamefont{Cava}},
  \bibinfo{journal}{Nature Physics} \textbf{\bibinfo{volume}{2}},
  \bibinfo{pages}{544} (\bibinfo{year}{2006}).
\end{thebibliography}
\end{document}

\title{Supplementary Material:\\Electronic correlations stabilizing time-reversal broken chiral superconductivity in single-trilayer TiSe$_2$}

\date{Version 4, 18 Nov 2013, compiled \today}

\author{R. Ganesh}
\affiliation{Institute for Theoretical Solid State Physics, IFW-Dresden, D-01171 Dresden, Germany }
\author{G. Baskaran}
\affiliation{The Institute of Mathematical Sciences, C.I.T. Campus, Chennai 600 113, India}
\affiliation{Perimeter Institute for Theoretical Physics, Waterloo, Ontario, Canada N2L 2Y5}
\author{Jeroen van den  Brink}
\affiliation{Institute for Theoretical Solid State Physics, IFW-Dresden, D-01171 Dresden, Germany }
\affiliation{Department of Physics, TU Dresden, D-01062 Dresden, Germany}
\author{Dmitry V. Efremov}
\affiliation{Institute for Theoretical Solid State Physics, IFW-Dresden, D-01171 Dresden, Germany }

\maketitle
 \date{\today}

\renewcommand{\theequation}{S\arabic{equation}}
\renewcommand{\thefigure}{S\arabic{figure}}
\renewcommand{\thetable}{S\arabic{table}}

\setcounter{equation}{0}
\setcounter{figure}{0}
\setcounter{table}{0}

%\section{Supplementary materials.}

\section{Band structure and allowed couplings}

\begin{figure}
\centering
\includegraphics[width=2.75in]{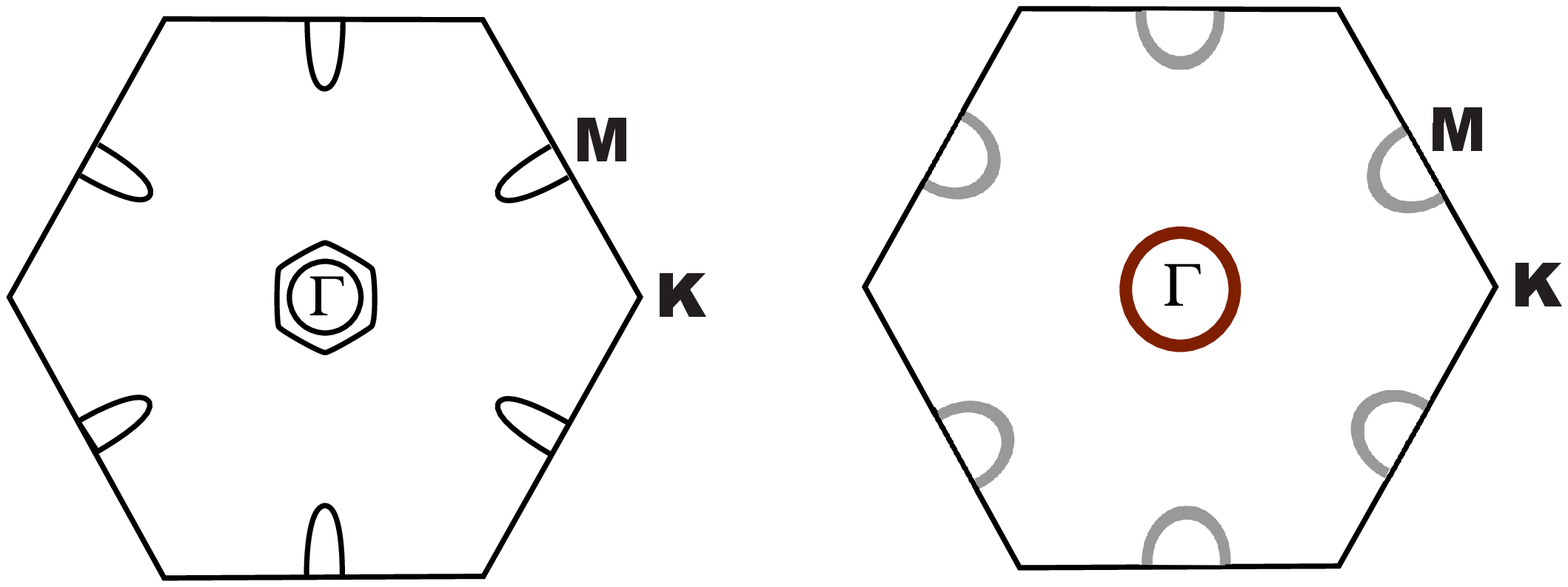}
\caption{Left: Fermi surfaces of two dimensional TiSe$_2$ slab obtained from {\it ab initio} calculations. Right: Approximate Fermi surface geometry considered for RG calculations. }
\label{fig.bandstruc}
\end{figure}

\begin{figure}
\centering
\includegraphics[width=3.25in]{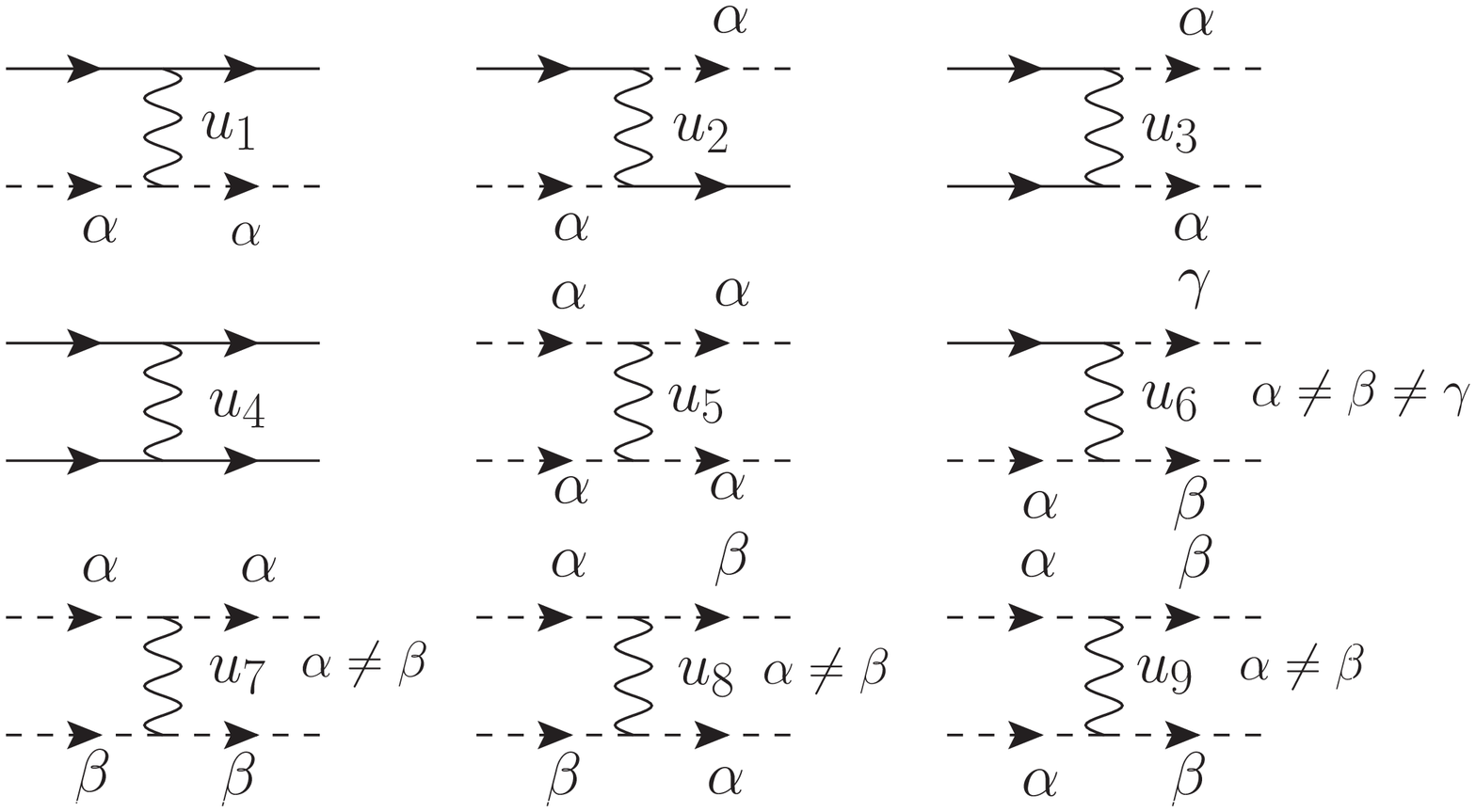}
\caption{The nine quartic couplings present in the low energy theory. Solid lines represent electrons that live on the central pocket. Dashed lines with coefficients $\alpha = 1,2,3$ represent electrons the three electron-like pockets at the Brillouin zone edge centres.}
\label{fig.barecouplings}
\end{figure}

The Fermi surfaces obtained from \textit{ab initio}
calculations are depicted in Fig.~\ref{fig.bandstruc}a. Fig.~\ref{fig.bandstruc}b shows the approximate Fermi surface structure that we have used in the RG calculations. 
This configuration allows for nine different interaction processes, shown in Fig.~\ref{fig.barecouplings}. These couplings are included in the Lagrangian defined in the main text.

\section{Parquet diagrams for RG flow}
One-loop RG proceeds by integrating out higher energy excitations over a floating energy scale. For a `renormalizable' system, this process preserves the structure of low energy scatterings allowing a systematic treatment. The conditions for renormalizability are met in our two dimensional system which has a constant density of states as we take our Fermi surfaces to be circular.
Assuming perfect nesting between electron and hole pockets, both particle-particle and particle-hole bubbles contribute in the RG process, leading to a logarithmic correction to low energy scatterings.

Going from 2D TiSe$_2$ to 3D TiSe$_2$, the quasi-circular bands elongate along the axis to form distorted cylinders. However, the 3D material has one extra band which crosses the Fermi level -- a spherical band around the $\Gamma$ point. This band ruins renormalizability by introducing a strong energy-dependence into the density of states at the Fermi level. However, we still expect our results to apply to 3D TiSe$_2$ where the central band may not play a crucial role.

\begin{figure}
\centering
\includegraphics[width=3.5in]{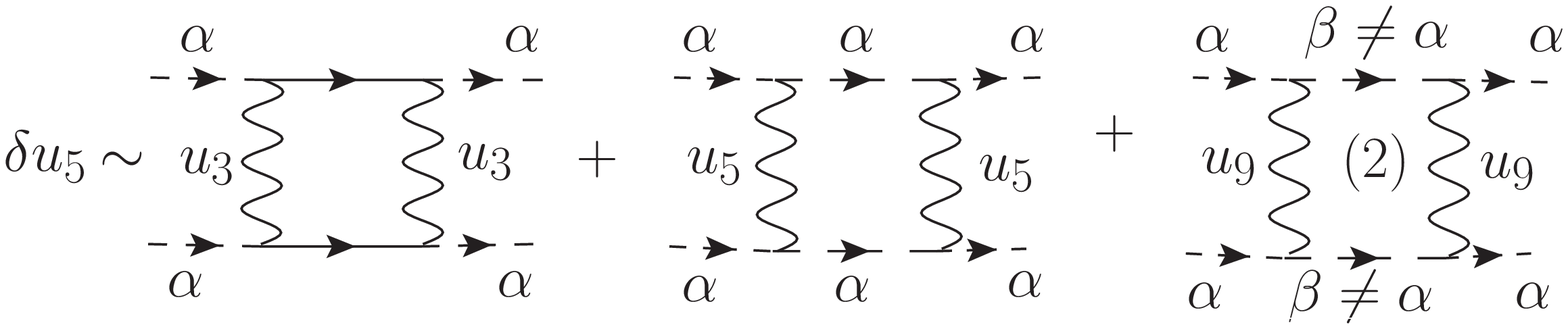}
\caption{Parquet diagrams for the RG flow of coupling $u_5$, representing inter-pocket scattering within the $\alpha$ pocket.  }
\label{fig.parquet}
\end{figure}

RG flow equations derived from parquet diagrams are given in Eqs.~2 in the main text. As an illustration, the derivation of the $u_5$ equation is shown diagrammatically in Fig.~\ref{fig.parquet}

\section{Gap equations and effective vertices}

\begin{figure}
\centering
\includegraphics[width=3.5in]{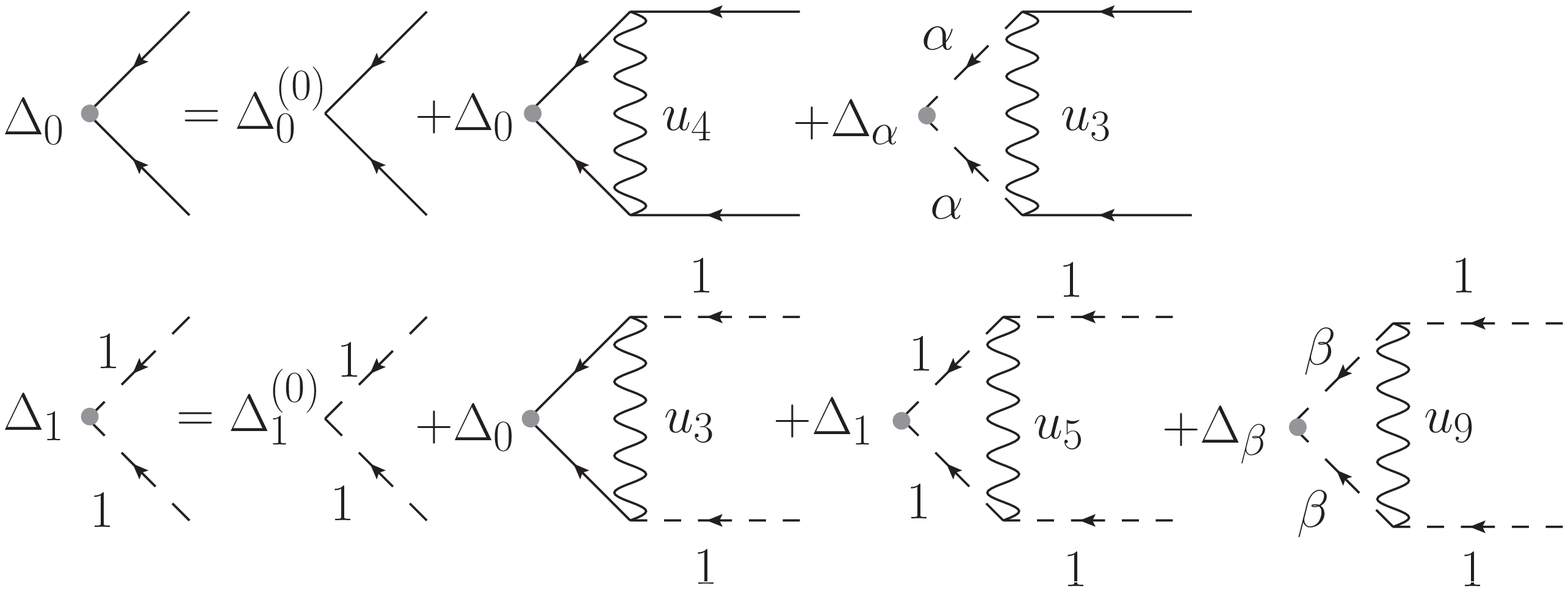}
\caption{Gap equations for superconducting orders. Top: dressing of a pairing field in the central pocket. The index $\alpha$ has to be summed over all three M pockets. Bottom: dressing of a pairing field acting on the $M_1$ pocket. The index $\beta$ has to be summed over $\{2, 3\}$, the other two M pockets.}
\label{fig.SCgapeq}
\end{figure}

To test SC susceptibility above $T_c$, we introduce external pairing fields $\Delta_\alpha^{(0)}$. As shown in Fig.~\ref{fig.SCgapeq}, they are dressed by interactions giving rise to response fields
\bea
\nonumber \left(\begin{array}{c}
\Delta_0 \\ \Delta_1 \\ \Delta_2 \\ \Delta_3
\end{array}\right) =
\left\lbrace I_{4\times4} + C
\left( \begin{array}{cccc}
u_4 & u_3 & u_3 & u_3 \\
u_3 & u_5 & u_9 & u_9 \\
u_3 & u_9 & u_5 & u_9 \\
u_3 & u_9 & u_9 & u_5
\end{array}\right)
\right\rbrace^{-1}
\left(\begin{array}{c}
\Delta_0^{(0)} \\ \Delta_1^{(0)} \\ \Delta_2^{(0)} \\ \Delta_3^{(0)}
\end{array}\right).
\eea
Here, $C = T \sum_{\omega_m} \!\!\int \!\frac{d^2k}{(2\pi)^2} \mathcal{G}^{0/\alpha}(\bk,\omega_m) \mathcal{G}^{0/\alpha}(-\bk,-\omega_m)$ represents the particle-particle or hole-hole bubble (their values are assumed to be equal due to nesting). The Green's functions $\mathcal{G}^{0}(\bk,\omega_m)$ and $\mathcal{G}^{\alpha}(\bk,\omega_m)$ are on the central pocket and on the $M_\alpha$ pocket ($\alpha=1,2,3$) respectively.
These equations are diagonalized by the transformation
\bea
\nonumber \left(\begin{array}{c}
\Delta_{chiral}^{clock.} \\ \Delta_{chiral}^{anti-cl.} \\ \Delta_{s_{++}} \\ \Delta_{s_{+-}}
\end{array}\right) =
\left( \begin{array}{cccc}
0 & 1 & e^{i2\pi/3} & e^{i4\pi/3} \\
0 & 1 & e^{i4\pi/3} & e^{i2\pi/3} \\
\lambda_+ & 1 & 1 & 1 \\
\lambda_- & 1 & 1 & 1
\end{array}\right)
\left(\begin{array}{c}
\Delta_0 \\ \Delta_1 \\ \Delta_2 \\ \Delta_3
\end{array}\right).
\eea
$\Delta_{chiral}^{clock.}$ and $\Delta_{chiral}^{anti-cl.}$ represent degenerate, clockwise and anti-clockwise, order parameters of chiral superconductivity. 

In the above equation, $\lambda_\pm = \{(u_4-u_5-2u_9)/2 \pm R \}/u_3$; where $R$ is as defined in the main text.  
Provided $u_3 > 0$, the $\Delta_{s_{+-}}$ solution satisfies ($sign(\Delta_M) = - sign(\Delta_\Gamma)$), corresponding to $s_{+-}$ pairing. The $\Delta_{s_{++}}$ solution also has the correct symmetry for $s_{++}$ order. However, if $u_3<0$, the characters of the solutions are swapped; the effective vertices of $s_{++}$ and $s_{+-}$ pairing are interchanged. However, the flow equation for $u_3$ is such that it can never change sign. As the bare Coulombic coupling is positive, $u_3$ always remains positive. 

In terms of these new SC order parameters, the dressed response fields simplify to
\bea
\nn \Delta_{chiral}^{clock./anti-cl.} &=& \{1 + C \cdot \Gamma^{SC}_{chiral} \}^{-1} \Delta_{chiral}^{clock./anti-cl. (0)}, \\
\nn \Delta_{s_{+-}} &=& \{1 + C \cdot \Gamma^{SC}_{s_{+-}} \}^{-1}
\Delta_{s_{+-}}^{(0)}, \\
\Delta_{s_{++}} &=& \{1 + C \cdot \Gamma^{SC}_{s_{++}}\}^{-1} \Delta_{s_{++}}^{(0)}.
\eea
Upon cooling to the critical temperature, provided $\Gamma^{SC}<0$, the quantity within braces will vanish; the response field can then be non-zero even when the external field is infinitesimal. This condition gives $T_c$. The order with the highest $\Gamma$ will have the highest $T_c$. 

\begin{figure}
\centering
\includegraphics[width=3.5in]{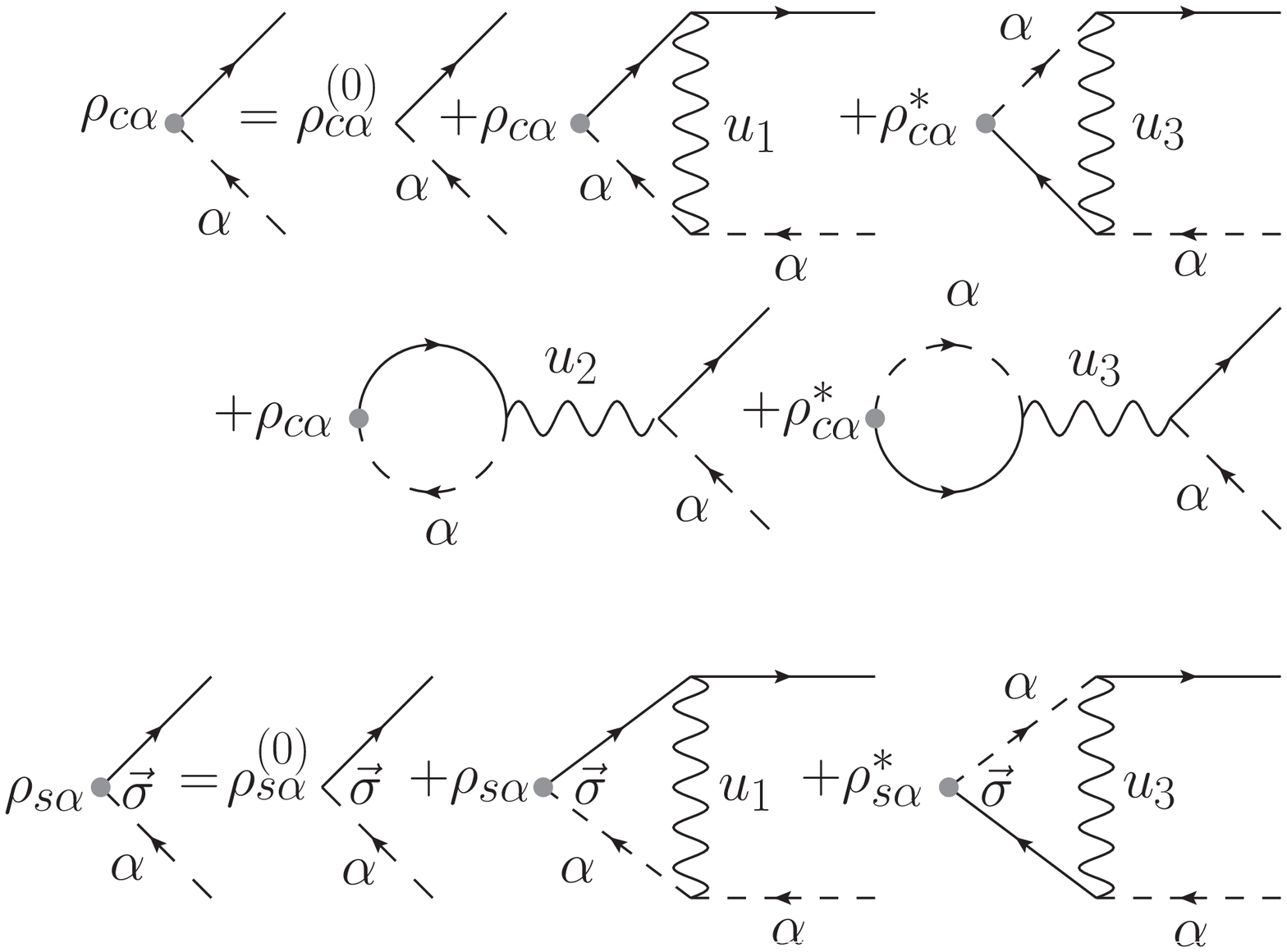}
\caption{Diagrammatic representation of the gap equations for CDW and SDW orders.}
\label{fig.gapeq}
\end{figure}

The effective vertices for CDW and SDW order can be derived similarly. The gap equations for CDW and SDW order are shown diagrammatically in Fig.~\ref{fig.gapeq}. 
These diagrams give linearly decoupled order parameters $(\rho_{c/s}^r)* =  (\rho_{c/s}^r)$ and  $(\rho_{c/s}^i)* = - (\rho_{c/s}^i)$ for both CDW and SDW. We will designate them as `real' and `imaginary' order parameters.
We extract the effective vertices from the linearized gap equation; i.e., approaching from high temperatures, we assume that ordering sets in at some transition temperature. This requires the following condition,
\bea
\nonumber 1\!\!\! &=&\!\!\! -T_{CDW}^{r,i}  \Gamma^{CDW}_{(r,i)} \sum_{\omega_m} \!\!\int \!\frac{d^2k}{(2\pi)^2} \mathcal{G}^{0}(\bk,\omega_m) \mathcal{G}^{\alpha}(\bk +M_i,\omega_m), \\
\nonumber 1\!\!\! &=&\!\!\! -T_{SDW}^{r,i} \Gamma^{SDW}_{(r,i)} \sum_{\omega_m} \!\!\int\! \frac{d^2k}{(2\pi)^2} \mathcal{G}^{0}(\bk,\omega_m) \mathcal{G}^{\alpha}(\bk +M_i,\omega_m),
\eea
where $r$ and $i$ denote real and imaginary parts of the order parameters.
$T_{SDW}^{i}$ and $T_{SDW}^{r}$ are transition temperatures for the order parameters $\rho_{s}^{r}$ and $\rho_{s}^{i}$ respectively. 
The effective vertices for CDW order are given $\Gamma^{CDW}_{real} = u_1-u_3-2u_2$ and $\Gamma^{CDW}_{imag} = u_1+u_3-2u_2$. The  vertices for SDW order are $\Gamma^{SDW}_{real} = u_1+u_3$ and $\Gamma^{SDW}_{imag} = u_1-u_3$.

\section{Role of the chemical potential}
\begin{figure}
\centering
\includegraphics[width=3.5in]{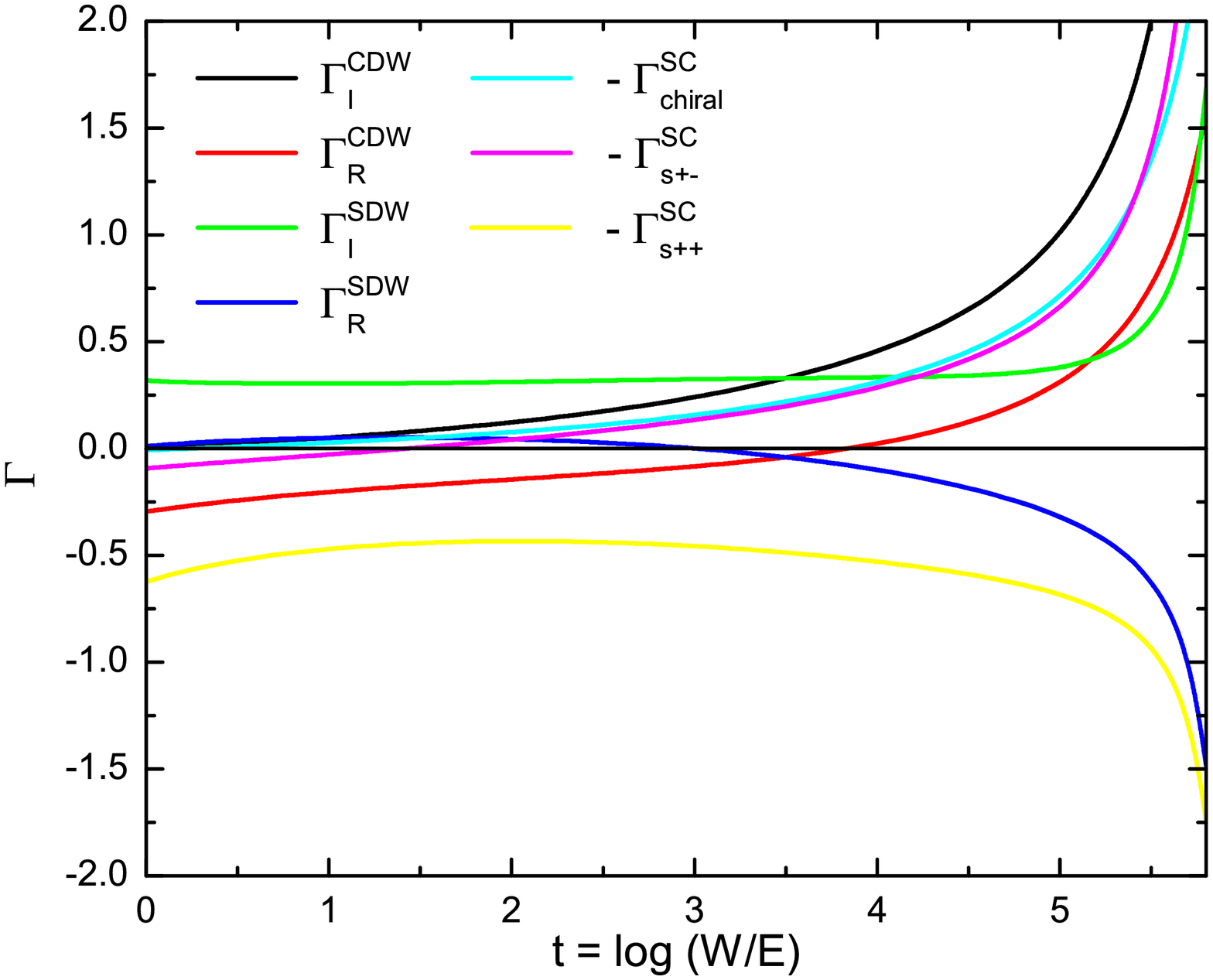}
\caption{Flow of effective vertices for bare interactions estimated as in main text, with screening parameter $\zeta=1.07$. The bare interactions are 
 $u^{(0)}_1/\zeta = u^{(0)}_2 = u^{(0)}_3=0.82u^{(0)}/\zeta $, $ u^{(0)}_4 = 1.85u^{(0)}$, $u^{(0)}_5/\zeta =u^{(0)}_7/\zeta=u^{(0)}_8=u^{(0)}_9= 0.6u^{(0)}/\zeta$ and $ u^{(0)}_6 = 0.675u^{(0)}/\zeta$, with $u^{(0)}=0.2$.
}
\label{fig.flowzeta107}
\end{figure}

At each step of RG flow, we integrate out excitations that lie between $E$ and $E+dE$, where $E$ is the floating RG scale. We track the resulting correction to low energy scattering processes that involve quasiparticles near the Fermi surface. As $E$ decreases from the bandwidth down to the Fermi energy, the flow of couplings separates into two regimes. At higher energies $E>\mu$ (we measure energies from the Fermi level),
when integrating out higher energy excitations, we can approximate the total incoming momentum to be zero. This RG group flow is captured by parquet diagrams. The particle-particle and particle-hole channels are coupled due to umklapp processes $U_3$ and $U_6$. The renormalization flow is captured by parquet diagrams leading to the Eqs.~2 in the main text. At low energies below $\mu$,
the polarization correction depends on the incoming momentum of the scattering process. The two channels decouple -- flow is be described by separate ladder diagrams for each channel\cite{Maiti2010}.
Thus, flow according to parquet diagrams can be cut off at $\mu$. For example, Fig.~\ref{fig.flowzeta107} shows RG flow which ultimately goes to the $s_{+-}$ SC fixed point. However, during most of the duration of RG flow, the effective vertex for imaginary CDW order dominates. When the floating RG scale reaches $\mu$, CDW order may have the largest effective vertex even though the ultimate fixed point corresponds to $s_{+-}$ SC order. Below $\mu$, the effective vertices for CDW and superconducting orders will evolve according to different sets of ladder diagrams. It is then possible that CDW order may set in at a higher temperature, thus pre-empting superconductivity. For large values of $\zeta \gtrsim 1.6$, we find that SDW order may pre-empt superconductivity. 

